\documentstyle[preprint,aps]{revtex}
\tightenlines
\def\dd{{\rm d}}\def\ee{{\rm e}}
\def\half{{\textstyle\frac12}}\def\fourth{{\textstyle\frac14}}

\begin{document}
\draft

\title{Expansion-induced contribution\\ to the precession of binary orbits}
\author{Brett Bolen, Luca Bombelli, and Raymond Puzio}
\address{Department of Physics and Astronomy, University of Mississippi,
\protect\\ 108 Lewis Hall, University, MS 38677}
\maketitle

\begin{abstract}
We point out the existence of new effects of global spacetime expansion on local
binary systems. In addition to a possible change of orbital size, there is a
contribution to the precession of elliptic orbits, to be added to the well-known
general relativistic effect in static spacetimes, and the eccentricity can change.
Our model calculations are done using geodesics in a McVittie metric, representing a
localized system in an asymptotically Robertson-Walker spacetime; we give a few
numerical estimates for that case, and briefly comment on ways in which the model
should be improved.
\end{abstract}


\noindent The issue of whether the global cosmological expansion affects local
gravitating systems, such as planetary systems and galaxies, has been studied for
a long time (for recent work, and discussions of previous results, see e.g.\
Refs.\ \cite{And,Bon,CFV,Nol2}). The specific questions asked in the literature
focus mostly on the extent to which local systems expand, and on the form and
magnitude of the corrections to the effective forces felt by orbiting test bodies;
while the approaches used vary, sometimes in conceptually important ways, the
consensus is that there is an effect in principle, but in practice it is
exceedingly, undetectably small. To a first approximation, a reasonable,
physically motivated point of view, as expressed by Misner, Thorne, and Wheeler
\cite{MTW}, for the case of galaxies, is that they are like rigid pennies attached
to the expanding balloon representing the universe, and they do not themselves
expand.

While we agree with this general conclusion, at least in terms of currently
feasible observations, an increase in orbit size is not the only effect the
cosmological expansion can have on local systems, nor the only cumulative one.
In the search for such effects described in this paper, we use as a model
spacetime an actual solution of Einstein's equation which represents a gravitating
body in an asymptotically expanding universe; restrict our attention to nearly
Newtonian orbits in the weak-gravity, slow-expansion approximation; and consider the
time variation of the parameters characterizing the corresponding Keplerian ellipses,
their size, perihelion angle, and eccentricity. As we will show, this approach leads
us to predict changes in all of those parameters as a consequence of global
expansion.

Before we proceed, however, we will make a few comments on the systems involved.
It is well known that Keplerian elliptical orbits in Newtonian gravity do not
precess (the $1/r$ potential is one of two that yield closed orbits, together
with the harmonic oscillator, as Bertrand's theorem states) but the
corresponding ones in general relativity do, while the size and eccentricity
of the general relativistic orbits are constant. These results hold for the
case of a single, spherically symmetric, isolated center of attraction, with
a test body and no other matter around it. Corrections due to other orbiting
bodies can be calculated, and they may affect all parameters of the orbits, but as
far as the global expansion is concerned, under the above conditions there can be
{\it no local effects\/} due to cosmological expansion in general relativity,
regardless of the rate at which the rest of the universe expands, because
Birkhoff's theorem then concludes that locally the spacetime must be the static,
Schwarzschild solution (Einstein and Straus \cite{Ein} also showed this explicitly
in one class of models, without using Birkhoff's theorem). The cosmological
expansion can be felt by a local system only if (i) The situation is not
spherically symmetric, and/or (ii) There is matter (some gas, dust, or dark
matter, say) within the local system, or a cosmological constant. This is one of
the many ways in which one can see that general relativity obeys the spirit of
Mach's principle, but only up to a point (see, e.g., Ref.\ \cite{Pac}); the global
behavior of distant objects can affect the orbital dynamics, but the dynamical
equations are local.

In a realistic model of a local system, both deviations from spherical
symmetry and extra matter will have an effect; in the solar system, for
example, the presence of other stars, solar multipoles such as oblateness,
minor objects, interplanetary matter and/or the solar wind and radiation, are
ways in which this happens. Studying the way the global expansion affects the
local system then means: (i) Understanding how the dynamics of surrounding
non-symmetric matter and/or the local internal environment of the system
depends on the global behavior of the universe, and (ii) Calculating the
effect of the local environment on the dynamics of the system of interest.
Notice that, for a local system that is considered as part of a larger one,
such as a planetary system inside a galaxy, the first part amounts to
assuming that a similar problem has been solved one level up; for example, the
spacetime metric for a planetary system should not be asymptotic to a
Robertson-Walker metric expanding at the Hubble rate, but to a metric that is
appropriate inside the galaxy, expanding, if at all, at a rate previously found
for that system. The problem thus becomes somewhat involved, but it seems to us
that this is the only way to avoid sneaking in an assumption about the local
environment's expansion as part of the setup, which has occasionally been done. On
the other hand, once it has been formulated in this way, the problem is broken down
into parts that can be separately investigated, and labelled by the structure scale
under consideration (galactic, planetary, ...), mechanism by which the effect may be
transmitted, and type of effect.

Here, as a start, we will limit ourselves to considering a simplified model to show
the effects that may arise. Spherical symmetry will be preserved, and the mechanism
responsible for transmitting environment effects will be the presence of matter in
the local system. In our numerical estimates we will use orbital parameters relevant
for a couple of known planets and galaxies, and the Hubble constant to obtain the
environment expansion rate; much more work will be needed on several aspects of this
problem, including the type of matter and spacetime model used and the role of
anisotropy, before we can claim to have a reasonable understanding of it along the
lines described above.

While spherically symmetric, vacuum, asymptotically flat spacetimes and
homogeneous, isotropic cosmological ones with fluid matter or a cosmological
constant can be easily treated in general relativity and give rise, respectively,
to the Schwarzschild solution and the Friedmann-Robertson-Walker or de~Sitter
spacetimes, solutions representing an isolated massive object embedded in an
expanding universe are more difficult to get, and fully explicit forms are not
usually given (see, for example, the general discussion in Ref.\ \cite{Sol}). The
known metrics representing isolated bodies in asymptotically FRW spacetimes are
mainly of three types: McVittie \cite{McV}, Einstein-Straus \cite{Ein}, and
Ferraris-Francaviglia-Spallicci \cite{FFS1}. In this paper, we will use McVittie
metrics, although an explicit solution is given in Ref.\ \cite{FFS1} where a
Schwarzschild and a FRW solution are matched smoothly across an extended region
of spacetime represented by a transition metric, where the orbit of a test
particle may also be analysed.

The (asymptotically) spatially flat McVittie model with $k = 0$, in the isotropic
coordinate form used by Hogan \cite{Hog}, has line element
\begin{equation}
    \label{metric1}
    \dd s^2 = -\bigg( \frac{1-M \ee^{-\beta/2}/2r}
    {1+M \ee^{-\beta/2}/2r} \bigg)^{\!2} \,\dd t^2
    + \ee^{\beta(t)}
    \bigg(1 + \frac{M\ee^{-\beta/2}}{2r} \bigg)^{\!4}
    \,(\dd r^2+r^2\dd\Omega^2) \;.
\end{equation}
Here, the constant $M$ is interpreted as the ``mass at the singularity," and
$\beta(t)$ represents the (asymptotic) expansion rate of the universe. We will not
give an explicit expression for $\beta(t)$, since it is related to the specific
equation of state of the fluid, but will think of it instead in terms of an
expansion $\beta(t) = \beta_0 + \dot\beta_0\,(t-t_0) + {\cal O}((t-t_0)^2)$,
around the present time $t_0$, say, where the first few coefficients can be fitted
phenomenologically.

The above line element has the advantage that, as $r \rightarrow \infty$, the $t$ =
const hypersurfaces become the surfaces of homogeneity of a Robertson-Walker model.
For our purposes however, we find it more convenient to work with a radial
coordinate with a geometrical meaning tied to the area of the corresponding
2-sphere, and use $R:= r\,\ee^{\beta/2} (1 + GM \ee^{-\beta/2}/2r)^2$ in the
interval $R\in(2M,\infty)$ \cite{Nol2}. We can then rewrite the line element as
\begin{equation}
    \label{metric2}
    \dd s^2
     = \bigg[-f(R) + \bigg(\frac{R\dot\beta}{2c}\bigg)^{\!2}\,\bigg] c^2\dd t^2
    + \frac{2}{\sqrt{f(R)}} \frac {R\dot\beta}{2c}\,c \dd t\, \dd R
    + \frac{\dd R^2}{f(R)} + R^2\,\dd\Omega^2 \,,
\end{equation}
where $f(R):= 1-2GM/(c^2 R)$, and we have restored all $c$'s and $G$'s.

The use of the McVittie metric as a model for studying local systems in FRW
universes has been the subject of criticism \cite{Sus,FFS2}, but in  a series of
papers \cite{Nol1,Nol2,Nol3} Nolan made a good argument for the $k = 0$ model by
studying its global properties in detail. Nolan also showed that, in an
appropriate sense, those solutions are the unique ones representing an
isolated mass surrounded by a shear-free perfect fluid in a spatially flat FRW
universe; the shear-free condition leads to special properties that we would not
expect a black hole to have \cite{Nol2}, but the metric does provide us with a
viable explicit model for our purposes.

Because of the spherical symmetry, we consider as usual our orbits to lie in the
$\sin\theta = 1$ plane, and drop $\theta$ from the whole treatment from now on. For
a particle of mass $m$ moving in the metric (\ref{metric2}), one way of deriving
the equations of motion is to use the Hamiltonian approach. The canonical momenta
obtained from the action $S[x] = (m/2) \int\dd\tau\,g_{\mu\nu}\,\dot x^\mu\dot
x^\nu$ are $p_\mu = m\, g_{\mu\nu}\dot x^\nu$, where the overdot on a dynamical
variable, $\dot t$, $\dot R$, or $\dot\phi$, denotes a derivative with respect to
proper time $\tau$ along a trajectory (whereas $\dot\beta$ just indicates the
derivative of the known function $\beta(t)$ with respect to its argument).

The test particle Hamiltonian ${\cal H} = (2m)^{-1}g^{\mu \nu}p_\mu p_\nu$ in
this case becomes
$$
    {\cal H} = \frac{1}{2m} \biggl(-\frac{p_t^2}{f(R)\,c^2}
    + \frac{2}{\sqrt{f(R)}}\frac{R \dot\beta}{2c}\,\frac{p_t p_R}{c}
    + \bigg[ f(R) - \Big(\frac{R\dot{\beta}}{2c} \Big)^2 \bigg]p_R^2
    + \frac{p_\phi^2}{R^2} \biggr)\; ,
$$
supplemented by the constraint that $g^{\mu \nu}p_\mu p_\nu = -(mc)^2$,
or ${\cal H} = -\half\,mc^2$.

Angular momentum is conserved, $p_\phi = L$, and as is common in celestial
mechanics orbit problems, we reexpress $R$ in terms of $u:= 1/R$ in the equations
of motion. If we perturbatively expand those equations in powers of $1/c$ and
$\dot\beta$, zeroth order terms correspond to a particle in Newtonian gravitation;
neglecting all $\dot\beta$ terms corresponds to doing calculations in a static
spacetime, and we recover the well-known Schwarzschild results; the expansion
effects we are interested in arise when leading order terms in $\dot\beta$ are kept.

The radial equation can be recast into an orbit equation in terms of $u(\phi)$,
$$
    u'' + u = \frac{GM m^2}{L^2} + \frac{3\,GM u^2}{c^2} -
    \frac{\dot\beta^2 m^2}{4 L^2 u^3} - \frac{GM\dot\beta m}{2 c^2 L u}\,u'\;.
$$
If we look for a solution of the form $u(\phi) = u_0 + \xi(\phi)$, where
$u_0 = GMm^2/L^2$ corresponds to a Newtonian circular orbit and $\xi(\phi)$ is a
small perturbation (see, e.g., Refs.\ \cite{RoN,OaR}), the linearized equation for
the deviation $\xi(\phi)$ is
\begin{equation}
    \label{dampedoscill}
    \xi''
    = \frac{3\,GMu_0^2}{c^2}
    - \frac{\dot\beta^2 m^2}{4 L^2 u_0^3}
    - \left(1-\frac{6GMu_0}{c^2}-\frac{3\dot\beta^2 m^2}{4 L^2 u_0^4}\right)\xi
    - \frac{GM\dot\beta m}{2 c^2 L u_0} \xi' \;.
\end{equation}

It should be noted that for a general McVittie metric, $\dot\beta =
\dot\beta(t(\phi))$, so, to simplify the solution of the equation, we will now
limit ourselves to considering the case $\ddot\beta = 0$; in terms of comoving
coordinates as in Eq.\ (\ref{metric1}), this is the linear expansion case, although
the metric for this model has a timelike Killing vector field, as can be seen from
the fact that the line element (\ref{metric2}) becomes time-independent (in fact,
it becomes equivalent to a Schwarzschild-de~Sitter solution \cite{Nol2}). Equation
(\ref{dampedoscill}) is then of the form $\xi^{\prime\prime} = f_{\rm c} -
\omega^2\,\xi + f_{\rm d}\,\xi'$, which can be viewed as that of an oscillator with
a constant force $f_{\rm c}$, modified by a $\dot\beta$ term which affects the
equilibrium displacement of the oscillator and thus the geodesic orbit size, but it
does {\it not\/} lead to orbit expansion (when $\ddot\beta = 0$); the frequency
$\omega$ is also modified by a $\dot\beta$ term, which contributes to the orbital
precession, and, interestingly, there is a damping term $f_{\rm d}$, which leads to
a change in orbital eccentricity. The equation admits a general solution of the form
$\xi(\phi) = {\cal C} + {\cal E}\ee^{{\cal B}\phi} \cos ({\cal A}\, \phi-\phi_0)$,
depending on arbitrary constants $\cal E$ and $\phi_0$, and where
$$
    {\cal A} := \sqrt{\omega^2-\fourth\,f_{\rm d}^2}
    \approx 1 - \frac{3GMu_0}{c^2} - \frac{3\dot\beta_0^2 m^2}{8 L^2 u_0^4}\;,
$$
$$
    {\cal B} := \half\,f_{\rm d}
    = -\frac{GM\dot\beta_0 m}{4c^2 L u_0}\;,
$$
$$
    {\cal C} := \frac{f_{\rm c}}{\omega^2}
    \approx \frac{3GMu_0^2}{c^2} - \frac{\dot\beta_0^2 m^2}{4 L^2 u_0^3}\;,
$$
from which, defining $u_0' = u_0+{\cal C}$ and $\varepsilon_0 = {\cal E}/(u_0+{\cal
C})$, we finally get for the inverse radial coordinate
\begin{equation}
    \label{u}
    u(\phi) = u_0' \big[1 + \varepsilon_0\,\ee^{{\cal B}\phi}
    \cos ({\cal A}\,\phi - \phi_0)\big] \;.
\end{equation}

The angle $\sigma$ by which the orbit (\ref{u}) precesses during each revolution can
be found from the fact that the change $\Delta\phi$ such that the argument of the
cosine increases by $2\pi$ is $2\pi+\sigma = 2\pi/{\cal A}$; i.e., for ${\cal
A}\approx 1$, $\sigma \approx 2\pi\,(1-{\cal A})$, or
\begin{equation}
   \label{sigma}
   \sigma \approx \frac{6\pi GM}{a(1-\varepsilon^2)\,c^2}
   + \frac{3\pi}{4}\,\frac{\dot\beta_0^2a^3(1-\varepsilon^2)^3}{GM}\;,
\end{equation}
where $\varepsilon = \varepsilon_0\,\ee^{{\cal B}\phi}$ is the orbital
eccentricity, and we have used the fact that for a Keplerian ellipse $1/u_0 =
a\,(1 - \varepsilon^2)$, with $a$ the semi-major axis. The first term in
(\ref{sigma}) is the Schwarzschild contribution, $\sigma_0$, as expected; the
second term is the expansion-induced one, and has the same sign as the first
term. The eccentricity also changes, by a fractional amount per revolution
$\Delta\varepsilon/\varepsilon = \ee^{2\pi{\cal B}} - 1$, or
\begin{equation}
   \label{epsilon}
   \frac{\Delta\varepsilon}{\varepsilon}
   \approx -\frac{\pi GM\dot\beta_0 m}{2c^2Lu_0}
   = -\frac{\pi\dot\beta_0}{2c^2}\,\sqrt{GM a\,(1-\varepsilon^2)}\;.
\end{equation}
Equations (\ref{sigma}) and (\ref{epsilon}) are our main results; they show that,
even in a model with no orbit expansion (this may just be a feature of the
$\ddot\beta = 0$ McVittie models used), global expansion has an effect on
periastron precession and eccentricity.

We can get a first crude estimate of the magnitude of these effects by evaluating
$\sigma$ and $\Delta\varepsilon / \varepsilon$ for a few known systems. For the
solar planets Mercury and Pluto, even using a Hubble parameter value $H_0$ around
75 km/s/Mpc for $\dot\beta_0$, which is certainly a vast overestimate, one obtains
for the relative size of the expansion-induced precession term in (\ref{sigma}),
values $\sigma_\beta / \sigma_0 \approx 3\times 10^{-21}$ and $7\times 10^{-13}$,
respectively, and for the relative eccentricity change in (\ref{epsilon}),
$\Delta\varepsilon/\varepsilon \approx -1.5\times 10^{-21}$/rev and $-1.5\times
10^{-20}$/rev, respectively. From the $M$ and $a$ dependence of these quantities
we see that none of the strong gravitational field situations in which other general
relativistic effects are studied, such as the binary pulsar \cite{Kar} or stellar
motion around the galactic center \cite{FrM} can give usably large values, and we
conclude that all expansion effects are negligible for the astrophysics of stellar
systems.

What about gravitationally bound pairs of galaxies? After all, as one might expect
from cosmological effects (and in contrast to the Schwarzschild ones), both
$\sigma_\beta / \sigma_0$ and $\Delta \varepsilon / \varepsilon$ grow with orbit
size $a$. We can get an idea if we apply equations (\ref{sigma}) and (\ref{epsilon})
to the Large Magellanic Cloud, at $a\approx50$ kpc from the Milky Way, with
$M\approx10^{11} \,M_\odot$, which gives $\sigma_\beta / \sigma_0 \approx 0.34$ and
$\Delta\varepsilon / \varepsilon \approx -8\times 10^{-11}$/rev, both significantly
larger than the previous values, by many orders of magnitude in the case of
$\sigma_\beta / \sigma_0$, in a regime where the use of $H_0$ for $\dot\beta$,
although probably still not appropriate, comes closer to being realistic.

So, if some of the predicted numbers are not necessarily small, are our expansion
effects cosmologically relevant? The problem here lies with the time scales involved
in galactic dynamics, which make any direct observation of time variations
impossible. Still, the fact that the Hubble distance-redshift relation shows local
deviations on small cosmological scales is an indication that on those scales local
interactions and global expansion effects coexist. Similar ideas have already
motivated, e.g., work on the effect of global expansion on the formation and
evolution of clusters of galaxies \cite{NaP}, and more recently on the effect of
inhomogeneities on the overall expansion \cite{Buc} or of a cosmological constant on
local dynamics \cite{Axe}, and one of the potential benefits of this line of research
is the possibility of gaining a new handle on dark matter. The present approach can
be seen as a model for the onset of this situation with simple, binary systems, and
as we have seen it can provide predictions for such systems. The model can be
improved to include more realistic sources, diffuse matter and anisotropy, as well
as possible multipole moments of the orbiting mass \cite{TaH}. It may then be
possible to identify statistical effects which can be measured by observing large
numbers of galaxy pairs, or early universe effects which left and imprint on later
evolution.

\bigskip\noindent{\bf Acknowledgement}

\bigskip\noindent We would like to thank Kumar Bhatt for many stimulating
conversations, without which this project might have never been started.

\def\ApJ{{\sl Astrophys. J.}}\def\CQG{{\sl Class. Quantum Grav.}}
\def\JMP{{\sl J. Math Phys.}}\def\MNRAS{{\sl Mon. Not. R. Astron. Soc.}}
\def\MPL{{\sl Mod. Phys. Lett.}}\def\PR{{\sl Phys. Rev.}}
\def\PRL{{\sl Phys. Rev. Lett.}}\def\RMP{{\sl Rev. Mod. Phys.}}

\end{document}